\documentstyle{article}
\begin{document}
\begin{flushright}
NUP-A-96-13 \\
\end{flushright}
\hspace{\fill}
\vspace{14mm}
\renewcommand{\thefootnote}{\fnsymbol{footnote}}
\begin{center}
\Large\bf{A U(1) Gauge Theory for \\ 
Antisymmetric Tensor Fields}
\end{center}

\hspace{\fill}
\vspace{3mm}

\begin{center}
{\large 
$\rm{Shinichi \; Deguchi}^{\,\it{a}}\!\!$
\footnote{E-mail: deguchi@phys.cst.nihon-u.ac.jp}, 
$\rm{Tadahito \; Nakajima}^{\,\it{b}}\!\!$
\footnote{E-mail: nakajima@phys.cst.nihon-u.ac.jp} and 
$\rm{Hideomi \; Totsuka}^{\,\it{c}}\!\!$
\footnote{E-mail: htotuka@phys.cst.nihon-u.ac.jp}}
\end{center}

\begin{center}
{\it ${}^{a)}$ 
Atomic Energy Research Institute, 
Nihon University, Tokyo 101} 
\end{center}
\begin{center}
{\it ${}^{b)}$ Department of Physics, Nihon University, Tokyo 101}
\end{center}
\begin{center}{\it ${}^{c)}$ Quantum Science, Nihon University, Tokyo 101}
\end{center}

\hspace{\fill}
\vspace{5mm}
\begin{center}
\bf{Abstract}
\end{center}

We show that a U(1) gauge theory defined in the configuration space for 
closed $p$-branes yields the gauge theory of a massless rank-$(p+1)$ 
antisymmetric tensor field and the Stueckelberg formalism for 
a massive vector field.   
\newpage
\renewcommand{\thefootnote}{\arabic{footnote}}

Antisymmetric tensor fields have been introduced in various theories. 
For instance, supergravity multiplets in many supergravity theories include 
antisymmetric tensor fields [1]. In $p$-brane theories it is known 
that rank-$(p+1)$ antisymmetric tensor fields couple naturally to $p$-branes 
[2].

In these theories, however, rank-$(p+1)$ $[\,p\geq1\,]$ antisymmetric tensor 
fields have not been understood in terms of differential geometry.  
In order to make geometric aspects of rank-$(p+1)$ antisymmetric tensor 
fields clear, Freund and Nepomechie have introduced the space of all closed 
$p$-manifolds embedded in space--time [3]. 
This space is nothing other than the configuration space for closed 
$p$-branes. Hereafter we refer to the space as  {\it closed p-manifold space}. 
Freund and Nepomechie pointed out that a rank-$(p+1)$ antisymmetric tensor 
field is geometrically characterized as a constrained connection on a U(1) 
bundle over the closed $p$-manifold space. In their paper, 
only the case $p=1$ was discussed in detail: they derived the gauge theory of 
a massless rank-two antisymmetric tensor field from a U(1) gauge theory in  
loop space.

In a recent paper [4], the U(1) gauge theory in loop space has been 
reconsidered more elegantly. Thereby it became possible to derive 
the Stueckelberg formalism for massive tensor fields, 
as well as the gauge theory of a massless rank-two antisymmetric tensor field, 
from the U(1) gauge theory in loop space [4,5].

The purpose of the present paper is to generalize the formulation in ref.[4] 
by replacing the loop space by the closed $p$-manifold space; 
we indeed obtain the gauge theory of a massless rank-$(p+1)$ antisymmetric 
tensor field from a U(1) gauge theory defined in the closed $p$-manifold 
space.  In addition we show that the U(1) gauge theory in closed $p$-manifold 
space yields the Stueckelberg formalism for a massive vector field.

We define a closed $p$-manifold space $\Omega^{p}M^{D}$ as the set of 
all closed $p$-manifolds embedded in $D$-dimensional Minkowski 
space $M^{D}$.   
An arbitrary closed $p$-manifold in $M^{D}$ 
with a parametrization $x^{\mu}=x^{\mu}(\vec{\sigma})$ 
[$\,\vec{\sigma}\equiv(\sigma^{1},\,\sigma^{2},\ldots,\sigma^{p})\,$] 
is represented as a point in $\Omega^{p}M^{D}$ denoted by 
coordinates $(x^{\mu\vec{\sigma}})$ with $x^{\mu\vec{\sigma}}
\equiv x^{\mu}(\vec{\sigma})$$\:$\footnotemark[1]{$^{)}$}.
\footnotetext[1]{$^{)}$ In this paper, 
the indices $\kappa$, $\lambda$, $\mu$, $\nu$ and $\xi$ take the values 
0, 1, 2, ..., $D-1$,  
while each of the indices $\vec{\rho}$, $\vec{\sigma}$, $\vec{\chi}$ and 
$\vec{\omega}$ is a set of $p$ real variables that take values 
necessary to parametrize a closed $p$-manifold.} 
The closed one-manifold space $\Omega^{1}M^{D}$ is nothing other than the loop 
space. In what follows, we treat the pair $(\mu\vec{\sigma})$ as a 
generalized index in $\Omega^{p}M^{D}$. 
Since each closed $p$-manifold in $M^{D}$ itself does not vary under 
reparametrizations, $x^{\mu}(\vec{\sigma})$ are reparametrization scalar 
functions on the $p$-manifold. Thus the transformation rule of the coordinates 
$(x^{\mu\vec{\sigma}})$ under an infinitesimal reparametrization 
$\vec{\sigma}\rightarrow\bar{\vec{\sigma}}=\vec{\sigma}
-\vec{\varepsilon}(\vec{\sigma})$ 
[$\,\vec{\varepsilon}\equiv(\varepsilon^{1},\,\varepsilon^{2},\ldots,
\varepsilon^{p})\,$] is determined to be 
\begin{eqnarray}
x^{\mu\vec{\sigma}}\rightarrow\bar{x}^{\mu\vec{\sigma}}
=x^{\mu\vec{\sigma}}+\varepsilon^{\alpha}(\vec{\sigma})
x'^{\mu}_{\alpha}(\vec{\sigma}) \;,
\end{eqnarray}
%
where $x'^{\mu}_{\alpha}(\vec{\sigma})\equiv
\partial x^{\mu}(\vec{\sigma})/\partial\sigma^{\alpha}$ 
$(\alpha=1,\,2,\ldots,p)$, 
and $\varepsilon^{\alpha}(\vec{\sigma})$ are infinitesimal functions of 
$\vec{\sigma}$.

Let us consider a U(1) gauge theory defined in $\Omega^{p}M^{D}$.  
The infinitesimal gauge transformation of a U(1) gauge field  
${\cal A}_{\mu\vec{\sigma}}[x]$ on $\Omega^{p}M^{D}$ is given by
\begin{eqnarray}
\delta{\cal A}_{\mu\vec{\sigma}}[x]=
\partial_{\mu\vec{\sigma}} \Lambda[x] \; ,  
\end{eqnarray}
%
where $\partial_{\mu\vec{\sigma}}\equiv\partial/\partial x^{\mu\vec{\sigma}}$, 
and $\Lambda$ is an infinitesimal scalar function on $\Omega^{p}M^D$.
Since the U(1) gauge symmetry has no relation with reparametrizations,  
$\Lambda$ has to be invariant under (1): 
\begin{eqnarray}
x'^{\mu}_{\alpha}(\vec{\sigma})\partial_{\mu\vec{\sigma}}\Lambda[x]=0  \;. 
\end{eqnarray}
%
Combination of (2) and (3) leads to the condition 
$\delta(x'^{\mu}_{\alpha}(\vec{\sigma}){\cal A}_{\mu\vec{\sigma}})=0$, 
which shows that $x'^{\mu}_{\alpha}(\vec{\sigma}){\cal A}_{\mu\vec{\sigma}}$ 
is gauge invariant. Hence  
$x'^{\mu}_{\alpha}(\vec{\sigma}){\cal A}_{\mu\vec{\sigma}}$ will be written 
in terms of transverse components of ${\cal A}_{\mu\vec{\sigma}}$. 
However, such an expression is incompatible with the fact that 
$x'^{\mu}_{\alpha}(\vec{\sigma}){\cal A}_{\mu\vec{\sigma}}$ is Lorentz scalar, 
since the transverse components are dependent on a choice of coordinate system. 
Consequently, we conclude that 
\begin{eqnarray}
x'^{\mu}_{\alpha}(\vec{\sigma}){\cal A}_{\mu\vec{\sigma}}[x]=0  \;.
\end{eqnarray}
%
The reparametrization-invariant condition for ${\cal A}_{\mu\vec{\sigma}}$ 
is found to be 
\begin{eqnarray}
x'^{\mu}_{\alpha}(\vec{\sigma})\partial_{\mu\vec{\sigma}}
{\cal A}_{\nu\vec{\rho}}+\delta'_{\alpha}(\vec{\sigma}-\vec{\rho\,})
{\cal A}_{\nu\vec{\sigma}}=0  \;,  
\end{eqnarray}
%
where $\delta(\vec{\sigma})\equiv\prod_{\alpha=1}^{p}\delta(\sigma^{\alpha})$ 
and $\delta'_{\alpha}(\vec{\sigma})\equiv
\partial\delta(\vec{\sigma})/\partial\sigma^{\alpha}$.

We now assume the existence of a reparametrization-invariant measure  
$[dx]=d^{D}x \{dy\}$ of $\Omega^{p}M^{D}$. 
Here $d^{D}x \equiv \prod_{\mu}dx^{\mu}$ is defined from 
$x^{\mu}\equiv\oint {{d^{p}\sigma}\over{\varpi}} x^{\mu}(\vec{\sigma})$ 
$(\,\varpi\equiv\oint d^{p}\sigma\,)$, which denote a point in $M^{D}$,  
and $\{dy\}$ is a measure defined from $y^{\mu}(\vec{\sigma})\equiv 
x^{\mu}(\vec{\sigma})-x^{\mu}$. 
In a similar fashion to the U(1) gauge theory in loop space [3,4], 
we insert the damping factor ${\rm exp}(-L/v^{2})$ with 
$L\equiv\oint {{d^{p}\sigma}\over{\varpi}} {\rm det}\{-\eta_{\mu\nu}
x'^{\mu}_{\alpha}(\vec{\sigma})x'^{\nu}_{\beta}(\vec{\sigma})\}$ 
into an action for ${\cal A}_{\mu\vec{\sigma}}$ so that it becomes 
well-defined. 
Here $v\:(>0)$ is a constant with dimensions of $[{\rm length}]^{p}$, and  
$\eta_{\mu\nu}$, ${\rm diag}\eta_{\mu\nu} 
=(1, \,-1, \,-1, \, ..., \,-1)$, is the metric tensor on $M^{D}$. 
The action for ${\cal A}_{\mu\vec{\sigma}}$ with the damping factor 
is given by
\begin{eqnarray}
S_{\rm R}={k\over{V_{\rm R}}} \int [dx] {\cal L} \,{\rm exp}
\!\left(-{L\over{v^{2}}} \right) \; \end{eqnarray}
%
with the Maxwell-type lagrangian
\begin{eqnarray}
{\cal L}=-{1\over4}{\cal G}^{\kappa\vec{\rho},\lambda\vec{\sigma}} 
{\cal G}^{\mu\vec{\chi},\nu\vec{\omega}} 
{\cal F}_{\kappa \vec{\rho},\mu \vec{\chi}} 
{\cal F}_{\lambda \vec{\sigma},\nu \vec{\omega}} \;, \:{\footnotemark[2]}^{)}
\end{eqnarray}
%
\footnotetext[2]{$^{)}$ We employ Einstein's convention for indices in   
$\Omega^{p} M^{D}$; for example,  
$V^{\mu\vec{\sigma}}W_{\mu\vec{\sigma}} = \sum_{\mu}
\oint {{d^{p}\sigma}\over{\varpi}} 
V^{\mu\vec{\sigma}}W_{\mu\vec{\sigma}}$.}where  
$k$ is a constant, $V_{\rm R}\equiv\int\{dy\}{\rm exp}(-L/v^{2})$, and 
\begin{eqnarray}
{\cal F}_{\mu\vec{\rho},\nu\vec{\sigma}}\equiv
\partial_{\mu\vec{\rho}}{\cal A}_{\nu\vec{\sigma}}
-\partial_{\nu\vec{\sigma}}{\cal A}_{\mu\vec{\rho}} \;.
\end{eqnarray}
%
The (inverse) metric tensor ${\cal G}^{\mu\vec{\rho},\nu\vec{\sigma}}$ 
can be constructed so as to incorporate reparametrization invariance into 
the lagrangian ${\cal L}$. (Such a metric was actually defined in the U(1) 
gauge theory in loop space [4].) 
In the present paper, however, we use the metric 
${\cal G}^{\mu\vec{\rho},\nu\vec{\sigma}}=
\eta^{\mu\nu}\delta(\vec{\rho}-\vec{\sigma})$, disregarding the  
reparametrization invariance of ${\cal L}$. 
We understand that the integrations with respect to $y^{0}(\vec{\sigma})$  
are carried out after applying the Wick rotation 
$y^{0}(\vec{\sigma})\rightarrow -iy^{0}(\vec{\sigma})$. 

Let us now try to derive the local gauge theory of a massless rank-$(p+1)$  
antisymmetric tensor field on $M^{D}$. 
The simplest solution of (3) is 
\begin{eqnarray}
\Lambda^{(0)}[x]\equiv
\oint {{d^{p}\sigma}\over{\varpi}} {q_{0}\over{p!}} 
\Sigma^{\mu_{1}\mu_{2}\cdots\mu_{p}}(\vec{\sigma})
\lambda_{\mu_{1}\mu_{2}\cdots\mu_{p}}(x(\vec{\sigma})) \;,   
\end{eqnarray}
%
where $\Sigma^{\mu_{1}\mu_{2}\cdots\mu_{p}}(\vec{\sigma})\equiv
{x'_{1}}\!^{[\mu_{1}}(\vec{\sigma}) x'^{\mu_{2}}_{2}(\vec{\sigma})\cdots
x'^{\mu_{p}]}_{p}(\vec{\sigma})$, 
$q_{0}$ is a constant with dimensions of ${\rm [length]}^{-p}$,   
and $\lambda_{\mu_{1}\mu_{2}\cdots\mu_{p}}$ is an infinitesimal 
antisymmetric tensor function on $M^{D}$. 
The solution of (4) associated with $\Lambda^{(0)}$ is given by  
\begin{eqnarray}
{\cal A}^{(0)}_{\mu\vec{\sigma}}[x]\equiv
{q_{0}\over{p!}}\Sigma^{\nu_{1}\nu_{2}\cdots\nu_{p}}(\vec{\sigma})
B_{\mu\nu_{1}\nu_{2}\cdots\nu_{p}}(x(\vec{\sigma})) \;, 
\end{eqnarray}
%
where $B_{\mu\nu_{1}\nu_{2}\cdots\nu_{p}}$ is a rank-$(p+1)$ antisymmetric 
tensor field on $M^{D}$. 
It is verified that ${\cal A}^{(0)}_{\mu\vec{\sigma}}$ satisfies the  
condition (5). 
Substitution of (9) and (10) into (2) yields the gauge transformation
\begin{eqnarray}
\delta B_{\mu\nu_{1}\nu_{2}\cdots\nu_{p}}(x)={1\over{p!}}
\partial_{[\mu}\lambda_{\nu_{1}\nu_{2}\cdots\nu_{p}]}(x) \:.
\end{eqnarray}
%
Substituting (10) into (8), we obtain the field strength of  
${\cal A}^{(0)}_{\mu\vec{\sigma}}$\,: 
\begin{eqnarray}
{\cal F}^{(0)}_{\mu\vec{\rho},\nu\vec{\sigma}}[x]
={q_{0}\over{p!}}\delta(\vec{\rho}-\vec{\sigma})
\Sigma^{\lambda_{1}\lambda_{2}\cdots\lambda_{p}}(\vec{\sigma})
F_{\mu\nu\lambda_{1}\lambda_{2}\cdots\lambda_{p}}(x(\vec{\sigma}))
\end{eqnarray}
%
with
\begin{eqnarray}
F_{\mu\nu\lambda_{1}\lambda_{2}\cdots\lambda_{p}}(x)
\equiv{1\over(p+1)!} 
\partial_{[\mu}B_{\nu\lambda_{1}\lambda_{2}\cdots\lambda_{p}]}(x) \;,
\end{eqnarray}
%
which is invariant under the transformation (11). 
The field strength of ${\cal A}^{(0)}_{\mu\vec{\sigma}}$ is thus written in 
terms of that of $B_{\mu\nu_{1}\nu_{2}\cdots\nu_{p}}$.  
This result makes easy to derive an action for  
$B_{\mu\nu_{1}\nu_{2}\cdots\nu_{p}}$ from (6).

The lagrangian (7) with (12) becomes
\begin{eqnarray}
{\cal L}^{(0)}&\!\!\!=\!\!\!&
-{{q^{2}_{0}}\over{4(p!)^{2}}} \delta(\vec{0})
\oint{{d^{p}\sigma}\over{\varpi}}
\Sigma^{\lambda_{1}\cdots\lambda_{p}}(\vec{\sigma})
\Sigma^{\kappa_{1}\cdots\kappa_{p}}(\vec{\sigma})
\nonumber \\
& & \qquad \qquad \qquad \qquad \times
F_{\lambda_{1}\cdots\lambda_{p}\mu\nu}(x(\vec{\sigma}))
{F_{\kappa_{1}\cdots\kappa_{p}}}^{\mu\nu}(x(\vec{\sigma})) \;.
\end{eqnarray}
%
Recalling $x^{\mu}(\vec{\sigma})=x^{\mu}+y^{\mu}(\vec{\sigma})$, 
we carry out the Taylor expansion of the $F(x(\vec{\sigma}))^{2}$-term around 
$(x^{\mu})$:
\begin{eqnarray}
& &F_{\lambda_{1}\cdots\lambda_{p}\mu\nu}(x(\vec{\sigma}))
{F_{\kappa_{1}\cdots\kappa_{p}}}^{\mu\nu}(x(\vec{\sigma})) \nonumber \\
& &\quad=F_{\lambda_{1}\cdots\lambda_{p}\mu\nu}(x)
{F_{\kappa_{1}\cdots\kappa_{p}}}^{\mu\nu}(x) \nonumber \\
& &\qquad +\sum_{k=1}^{\infty} {1\over{k!}} \!\left( 
\prod_{j=1}^{k} \sum_{\mu_{j}=0}^{D-1} y^{\mu_{j}}(\vec{\sigma}) 
\partial_{\mu_{j}} \right)\! 
F_{\lambda_{1}\cdots\lambda_{p}\mu\nu}(x)
{F_{\kappa_{1}\cdots\kappa_{p}}}^{\mu\nu}(x)  \;.
\end{eqnarray}
%
All the differential coefficients at $(x^{\mu})$ in this Taylor series vanish  after integration  with respect to $x^{\mu}$,  
since $\mid\!x^{\mu}\!\mid<\infty$. 
Thus the action (6) with the lagrangian (14) is written as 
\begin{eqnarray}
S^{(0)}_{\rm R} &\!\!\!=\!\!\!&
-{{kq_{0}^{2}}\over{4(p!)^{2}}}\delta(\vec{0})
\int d^{D}x 
F_{\lambda_{1}\cdots\lambda_{p}\mu\nu}(x)
{F_{\kappa_{1}\cdots\kappa_{p}}}^{\mu\nu}(x) 
\nonumber \\ 
& & \times {1\over{V_{\rm R}}} \int \{dy\}
\oint{{d^{p}\sigma}\over{\varpi}}
\Sigma^{\lambda_{1}\cdots\lambda_{p}}(\vec{\sigma})
\Sigma^{\kappa_{1}\cdots\kappa_{p}}(\vec{\sigma})
{\rm exp} \!\left( -{L\over{v^{2}}} \right) 
\nonumber \\
&\!\!\!=\!\!\!&
{{(-1)^{p}(D-p)!}\over{4D!}} 
{{\partial\,{\rm log}V_{\rm R}}\over{\partial(1/v^{2})}}
kq_{0}^{2}\delta(\vec{0})
\int d^{D}x 
F_{\lambda_{1}\cdots\lambda_{p}\mu\nu}(x)
F^{\lambda_{1}\cdots\lambda_{p}\mu\nu}(x) 
\end{eqnarray}
%
by using the formula
\begin{eqnarray}
& &{1\over{V_{\rm R}}} \int \{dy\}
\oint{{d^{p}\sigma}\over{\varpi}}
\Sigma^{\lambda_{1}\cdots\lambda_{p}}(\vec{\sigma})
\Sigma^{\kappa_{1}\cdots\kappa_{p}}(\vec{\sigma})
{\rm exp} \!\left( -{L\over{v^{2}}} \right) 
\nonumber \\
&\!\!\!=\!\!\!& {{(-1)^{p+1}p!(D-p)!}\over{D!}} 
{{\partial\,{\rm log}V_{\rm R}}\over{\partial(1/v^{2})}}
\sum_{P}{\rm sgn}(P)\eta^{\lambda_{1}\kappa_{P(1)}}\cdots
\eta^{\lambda_{p}\kappa_{P(p)}} \;.
\end{eqnarray}
%
Here $P$ denotes a permutation of the numbers $1,\,2,\ldots,p$, and 
${\rm sgn}(P)$ takes $1(-1)$ for an even(odd) permutation. The summation 
$\sum_{P}$ extends over all possible permutations. 
Defining the constants $k$ and $q_{0}$ so as to satisfy the normalization 
condition 
\begin{eqnarray}
-{{(p+2)!(D-p)!}\over{2D!}} 
{{\partial\,{\rm log}V_{\rm R}}\over{\partial(1/v^{2})}}
kq_{0}^{2}\delta(\vec{0})=1 \;,
\end{eqnarray}
%
we arrive at the well-known action [1,2,6]
\begin{eqnarray}
S^{(0)}_{\rm R} = {{(-1)^{p+1}}\over{2(p+2)!}}\int d^{D}x 
F_{\lambda_{1}\cdots\lambda_{p}\mu\nu}(x)
F^{\lambda_{1}\cdots\lambda_{p}\mu\nu}(x) \;.
\end{eqnarray}
%
The condition (18) is understood with suitable regularizations 
of the divergent quantities.  
As a consequence, the U(1) gauge theory in $\Omega^{p}M^{D}$ leads to the 
gauge theory defined by (19).

Since ${\cal A}_{\mu\vec{\sigma}}$ is a functional field on $M^{D}$, 
it appears that ${\cal A}_{\mu\vec{\sigma}}$ contains an infinite number of 
local component fields. 
In fact, in addition to ${\cal A}^{(0)}_{\mu\vec{\sigma}}$\,, 
equation (4) has an infinite number of 
fundamental solutions written in terms of local fields on $M^{D}$. 
Among them, we next examine a fundamental solution consisting of vector and 
scalar fields on $M^{D}$.

Consider the following solution of (3):
\begin{eqnarray}
\Lambda^{(1)}[x]\equiv
\oint {{d^{p}\sigma}\over{\varpi}} 
q_{1}\sqrt{-\Sigma^{2}(\vec{\sigma})}\,\lambda(x(\vec{\sigma})) \;,
\end{eqnarray}
%
where $\Sigma^{2}(\vec{\sigma})\equiv
\Sigma_{\mu_{1}\mu_{2}\cdots\mu_{p}}(\vec{\sigma})
\Sigma^{\mu_{1}\mu_{2}\cdots\mu_{p}}(\vec{\sigma})$, 
$q_{1}$ is a constant with dimensions of ${\rm [length]}^{-p}$, 
and $\lambda$ is an infinitesimal scalar function on $M^{D}$. 
The solution of (4) associated with $\Lambda^{(1)}$ consists of a vector field
 $A_{\mu}$ and a scalar field $\phi$ on $M^{D}$: 
\begin{eqnarray}
{\cal A}^{(1)}_{\mu\vec{\sigma}}[x]\equiv
q_{1}\sqrt{-\Sigma^{2}(\vec{\sigma})} \left( {\delta_{\mu}}^{\nu}
+pQ_{\mu\lambda_{1}\cdots\lambda_{p-1}}(\vec{\sigma})
Q^{\nu\lambda_{1}\cdots\lambda_{p-1}}(\vec{\sigma}) \right)\!
A_{\nu}(x(\vec{\sigma}))   
\nonumber \\
+epx'^{\lambda_{1}}_{[1}(\vec{\sigma})\cdots 
x'^{\lambda_{p-1}}_{p-1}(\vec{\sigma})
\partial_{p]}Q_{\lambda_{1}\cdots\lambda_{p-1}\mu}(\vec{\sigma})
\phi(x(\vec{\sigma})) \;,
\end{eqnarray}
%
where $Q_{\mu_{1}\mu_{2}\cdots\mu_{p}}(\vec{\sigma})\equiv
\Sigma_{\mu_{1}\mu_{2}\cdots\mu_{p}}(\vec{\sigma})/\sqrt{-\Sigma^{2}
(\vec{\sigma})}\,$, 
$\partial_{\alpha}\equiv\partial/\partial\sigma^{\alpha}$, 
and $e$ is a constant with dimensions of ${\rm [length]}^{-(p-1)}$. 
Note that ${\cal A}^{(1)}_{\mu\vec{\sigma}}$ satisfies the condition (5). 
From (2) with (20) and (21), we have the gauge transformation 
\begin{eqnarray}
\delta A_{\mu}(x)=\partial_{\mu} \lambda(x)\;, \qquad
\delta \phi(x)={q_{1}\over e}\lambda(x) \;.
\end{eqnarray}
%
By direct calculation, the field strength of  
${\cal A}^{(1)}_{\mu\vec{\sigma}}$ is obtained as follows: 
\begin{eqnarray}
{\cal F}^{(1)}_{\mu\rho,\nu\sigma}[x]
&\!\!\!=\!\!\!&
q_{1}\delta(\vec{\rho}-\vec{\sigma}) 
\Bigl\{ 
\sqrt{-\Sigma^{2}(\vec{\sigma})} 
\,F_{\mu\nu}(x(\vec{\sigma})) 
\Bigr.
\nonumber \\ 
& & 
-p\Sigma^{\lambda_{1}\cdots\lambda_{p-1}\kappa}(\vec{\sigma})
Q_{\lambda_{1}\cdots\lambda_{p-1}[\mu}(\vec{\sigma})
F_{\nu]\kappa}(x(\vec{\sigma})) 
\nonumber \\ 
& & 
\Bigl.
+p x'^{\lambda_{1}}_{[1}(\vec{\sigma})\cdots 
x'^{\lambda_{p-1}}_{p-1}(\vec{\sigma})
\partial_{p]}Q_{\lambda_{1}\cdots\lambda_{p-1}[\mu}(\vec{\sigma})
\tilde{A}_{\nu]}(x(\vec{\sigma})) 
\Bigr\} 
\nonumber \\ 
& & 
+q_{1}p x'^{\kappa_{1}}_{[1}(\vec{\rho}^{\,})\cdots 
x'^{\kappa_{p-1}}_{p-1}(\vec{\rho}^{\,})
\delta'_{p]}(\vec{\rho}-\vec{\sigma})
\Bigl\{ 
(p-1)Q_{\mu\nu\kappa_{1}\cdots\kappa_{p-2}}(\vec{\rho}^{\,})
{\delta_{\kappa_{p-1}}}^{\xi}
\Bigr.
\nonumber \\ 
& & 
\Bigl.
-\Pi_{\mu\lambda_{1}\cdots\lambda_{p-1}\nu\kappa_{1}\cdots\kappa_{p-1}}
(\vec{\rho}^{\,})Q^{\lambda_{1}\cdots\lambda_{p-1}\xi}(\vec{\rho}^{\,}) 
\Bigr\}
\tilde{A}_{\xi}(x(\vec{\rho}^{\,}))
\nonumber \\
& &
-q_{1}p x'^{\kappa_{1}}_{[1}(\vec{\sigma})\cdots 
x'^{\kappa_{p-1}}_{p-1}(\vec{\sigma})
\delta'_{p]}(\vec{\sigma}-\vec{\rho}^{\,})
\Bigl\{ 
(p-1)Q_{\nu\mu\kappa_{1}\cdots\kappa_{p-2}}(\vec{\sigma})
{\delta_{\kappa_{p-1}}}^{\xi}
\Bigr.
\nonumber \\ 
& & 
\Bigl.
-\Pi_{\nu\lambda_{1}\cdots\lambda_{p-1}\mu\kappa_{1}\cdots\kappa_{p-1}}
(\vec{\sigma})Q^{\lambda_{1}\cdots\lambda_{p-1}\xi}(\vec{\sigma}) 
\Bigr\}
\tilde{A}_{\xi}(x(\vec{\sigma})) \;,
\end{eqnarray}
%
where
\begin{eqnarray}
{\Pi_{\mu_{1}\mu_{2}\cdots\mu_{p}}}^{\nu_{1}\nu_{2}\cdots\nu_{p}}(\vec{\sigma})
&\!\!\!\equiv\!\!\!&
{1\over(p-1)!}\,{\delta_{\mu_{1}}}^{[\nu_{1}} {\delta_{\mu_{2}}}^{\nu_{2}}
\cdots {\delta_{\mu_{p}}}^{\nu_{p}]}
\nonumber \\
& &
+pQ_{\mu_{1}\mu_{2}\cdots\mu_{p}}(\vec{\sigma})
Q^{\nu_{1}\nu_{2}\cdots\nu_{p}}(\vec{\sigma}) \;,
\end{eqnarray}
%
$F_{\mu\nu}\equiv\partial_{\mu}A_{\nu}-\partial_{\nu}A_{\mu}$ and 
$\tilde A_{\mu}\equiv A_{\mu}-(e/q_{1})\partial_{\mu}\phi$. 
Obviously the right-hand side of (23) is gauge invariant.

Substituting (23) into (6) and following the procedure for deriving (19), 
we finally arrive at the action in the Stueckelberg formalism: 
\begin{eqnarray}
S_{\rm R}^{(1)}=\int d^{D}x \left[ -{1\over4} F_{\mu\nu}(x)F^{\mu\nu}(x)
+{1\over2}m^{2} \tilde{A}_{\mu}(x)\tilde{A}^{\mu}(x) \right] \;.
\end{eqnarray}
%
In deriving this action, we have imposed a certain normalization condition on 
the coefficient of the $F(x)^{2}$-term. Then we find that the mass $m$ is 
proportional to $v^{-1/p}$.

In conclusion,  the gauge theory of a massless rank-$(p+1)$ antisymmetric 
tensor field and the Stueckelberg formalism for a massive vector field 
have been derived from the U(1) gauge theory in $p$-manifold space. 
In terms of differential geometry, ${\cal A}_{\mu\vec{\sigma}}$ is a  
connection on a U(1) bundle over $\Omega^{p}M^{D}$; as seen from (10) 
and (21), $B_{\mu\nu_{1}\nu_{2}\cdots\nu_{p}}$ and the pair of $A_{\mu}$ and 
$\phi$ are regarded as constrained connections on this U(1) bundle.

Recently, couplings of strings and massive local fields have been 
discussed from the point of view of the Lorentz force in loop space [7]. 
In a similar manner, we can consider couplings of $p$-branes and massive 
local fields such as $\tilde{A}_{\mu}$.

A Yang--Mills theory in loop space has also been studied. 
An essential idea in this theory is to take a loop group or an affine Lie 
group as the gauge group. It was shown that the Yang--Mills theory with the 
loop gauge group leads to a non-abelian Stueckelberg formalism for massive 
second-rank tensor fields [8], and that the Yang--Mills theory with the 
affine gauge group yields the Chapline--Manton coupling [9]. 
As a next study, it will be interesting to generalize the Yang--Mills theory 
in loop space by replacing the loop space by the $p$-manifold space. 
For the case $p=3$, a candidate for the gauge group will be 
the Mickelsson--Faddeev group [10]. This subject will be discussed in 
the future. 

\vspace{1.2cm}

We are grateful to Professor S. Naka and other members of the Theoretical  
Physics Group at Nihon University for their encouragements and useful 
comments. This work was supported in part by Nihon University Research Grant.

\newpage

\begin{center}
{\Large\bf References}

\vspace{1mm}

\end{center}
\begin{enumerate}
\item A. Salam and E. Sezgin, {\it Supergravities in Diverse Dimensions}, 
Vol. 1 and Vol. 2 (North-Holland, Amsterdam/World Scientific, Singapore, 
1989). 

\vspace{1mm}

A. H. Chamseddine, Phys. Lett. {\bf B 367} (1996), 134. 
%
\item M. J. Duff, R. R. Khuri and J. X. Lu, Phys. Rep. {\bf B259} (1995), 
213.  
\vspace{1mm}

Y. Ne'eman and E. Eizenberg, {\it Membranes and Other Extendons (p-branes)} 
(World Scientific, Singapore, 1995).  

\vspace{1mm}

S. Ishikawa, Y. Iwama, T. Miyazaki, K. Yamamoto, M. Yamanobe and R. Yoshida, 
Prog. Theor. Phys. {\bf 96} (1996), 227. 

\item P. G. O. Freund and R. I. Nepomechie, Nucl. Phys. {\bf B199} (1982), 
482. 
%
\item S. Deguchi and T. Nakajima, Int. J. Mod. Phys. {\bf A9} (1994), 1889. 
%
\item S. Deguchi and T. Nakajima, Prog. Theor. Phys. {\bf 94} (1995), 305. 
%
\item T. Kimura, Prog. Theor. Phys. {\bf 65} (1981), 338.
%
\item S. Deguchi, Phys. Rev. {\bf D52} (1995), 1284.  
%
\item S. Deguchi and T. Nakajima, Int. J. Mod. Phys. {\bf A10} (1995), 1019. 
%
\item S. Deguchi and T. Nakajima, hep-th/9606178, NUP-A-96-6.
%
\item L. D. Faddeev, Phys. Lett. {\bf B145} (1984), 81. 

\vspace{1mm}

J. Mickelsson, {\it Current Algebras and Groups} (Plenum Press, New York, 
1989). 

\vspace{1mm}

A. Pressley and G. Segal, {\it Loop Groups} (Oxford Univ. Press, New York, 
1986).

\end{enumerate}

\end{document}